\documentclass[a4paper]{spie}  %>>> use this instead for A4 paper
\newcommand{\binomial}[2]{\left(\begin{array}{c} #1\\ #2\end{array}\right)}

\usepackage[pdftex]{graphicx}
\pdfoutput=1

\title{Studies in the physics of evolution: creation, formation, destruction} 

\author{Rudolf Hanel\supit{a},  Peter Klimek\supit{a} and Stefan Thurner\supit{a,b}
\skiplinehalf
\supit{a}Complex Systems Research Group, HNO, Medical University of Vienna, 
W\"ahringer G\"urtel 18-20, A-1090, Vienna, Austria \\
\supit{b}Santa Fe Institute, 1399 Hyde Park Road, Santa Fe, NM 87501, USA
}

\begin{document} 
  \maketitle 

%%%%%%%%%%%%%%%%%%%%%%%%%%%%%%%%%%%%%%%%%%%%%%%%%%%%%%%%%%%%% 
\begin{abstract}
The concept of (auto)catalytic systems has become a cornerstone in understanding  
evolutionary processes in various fields. The common ground is the observation that for the production of new 
species/goods/ideas/elements etc. the pre-existence of specific other elements is a necessary condition. 
In previous work some of us showed that the dynamics of the catalytic network equation can be understood 
in terms of topological recurrence relations paving a path towards the analytic tractability of notoriously high dimensional 
evolution equations. We apply this philosophy to studies in socio-physics, bio-diversity and massive events of creation and destruction in technological and biological networks. Cascading events, triggered by small exogenous fluctuations, lead to dynamics strongly resembling the qualitative picture of Schumpeterian economic evolution. Further we show that this new methodology allows to mathematically treat a variant of the threshold voter-model of opinion formation on networks. For fixed topology we find distinct phases of mixed opinions and consensus. 
\end{abstract}


%%%%%%%%%%%%%%%%%%%%%%%%%%%%%%%%%%%%%%%%%%%%%%%%%%%%%%%%%%%%%
\section{Introduction}

This paper deals with the understanding of dynamics of systems capable of evolution. 
In particular we are interested in systems of technological evolution with biological evolution as a special case.
We are dealing with systems of elements (e.g. molecules, species, goods, thoughts) that do not only 
exist or decay, but have the possibility to form new elements through recombination of already existing elements. 

For example consider a soup of molecules. Some molecules in the soup can react chemically with each other and form new molecules. Some reactions will lead to products that are already existing in the soup. Other reactions may produce molecules that are new and therefore transform the soup-environment into something new.  This in return opens the possibility for chemical reactions that have not been possible before the existence of the new product. 
In this context the concept of the {\em adjacent possible}\cite{origin} or {\em adjacent probable}\cite{arthur2004} has been introduced. This is the set of objects that can possibly become produced within a 
given time span into the future. What {\em can} be produced in the next timestep depends crucially on the 
details of the set of elements that exist at a given time. 
Existing elements undergo an evaluation within the newly formed context\cite{arthur2004}.

In this paper we are basically interested in three types of phenomena. 
First, cascades of creation, second, cascades of destruction. The
existence of these kind of dynamics has been postulated in the Schumpeterian 
view on economic evolution\cite{schumpeter11} long time ago.
Third, we study {\em formation} processes in evolutionary dynamics. In this scenario elements exist in two different states 0/1
(like existing/not existing, on/off, active/inactive). Each element is connected to a set of other elements through a (fixed) interaction matrix. An element can change its state depending on the states of the elements it is connected to. For example, some type of molecule may change its state from non existing to existing when sufficiently many other molecules also exist -- or vice versa. This problem can be cast onto an opinion formation\cite{Axelrod97,Krapivsky03, Mobilia03, Castellano05, Castellano06,Lambetal07, Lambiotte07} scenario\cite{Lambiotte06}.

One of the substantial problems in dealing with evolutionary systems  is the notoriously high or even unbound 
dimensionality of systems undergoing evolution.
Just think of how many biological species exist on this planet  or how many different goods and services are available, e.g. in Europe. Every element (good/species/chemical substance) $i$ is present in an evolving system with some relative frequency $x_i$. $x_i$ is the normalized abundance of the element $i$, such that $\sum x_i=1$. Think of each $x_i$ as an entry  in a $d$-dimensional vector containing all possible elements in the universe. Inter-dependence between pairs of elements 
can be coded in a large {\em interaction matrix} $\alpha$. If, for instance, $i$ uses a service provided by $j$ we set $\alpha_{ij}=1$ otherwise $\alpha_{ij}=0$. For ternary (and higher) relations $\alpha$ is not a matrix but a tensor which we call the {\em production rule table} or simply {\em rule table}. Suppose we look at chemical reactions then $\alpha_{ijk}=1$ when $i$ is the product of a chemical reaction of the substrates $j$ and $k$, and $\alpha_{ijk}=0$ if this chemical process is not permitted by the rules of chemistry.

High dimensionality usually makes such systems particularly difficult to treat in an analytic fashion. On the other hand the combinatorial explosion of elements that emerges through the recombinations of other elements, severely limits 
computational analysis as well. 
The exact way one element influences another is often only poorly known, or not known at all.
For example, we do not exactly know how many biological species are living on this planet right now (although we may make educated guesses). Needless to say that we know even less about the actual interactions and relations that species involve in.
Consequently we usually can not exactly describe the inter-dependencies between species (e.g. trophic dependencies)  which  govern their dynamics. We are certainly incapable of prestating future inter-dependencies of species that might come into being in the future course of earth history. This is to say, that we usually know very little about the true structure of the interaction  matrices or tensors $\alpha$. However, we may happen to know several things, constraining our ignorance. For example, we might 
have knowledge about the average number of chemical reactions producing the same chemical substance, or the average number of trophic dependencies an individual in an ecosystem is maintaining. 

It was suggested\cite{hanel05} to meet the above problems of dimensionality and ignorance  by two model assumptions: 
(i) The focus is shifted from the actual frequency of elements $x_i$, to the system's diversity, where diversity is defined as the number of existing elements. An element exists, if $x_i>0$, and does not exist if $x_i=0$.    
(ii) The entries  of the interaction matrices or rule tables $\alpha$ that encode the relation between the specific elements of the system, are modelled as random. For simplicity the entries of $\alpha$ are 0 or 1.

These two assumptions are powerful tools in mapping the dynamics -- that ideally should be described on a level of differential equations of the relative frequencies (concentrations) of elements -- onto recurrence relations for the system diversity or other variables of interest.

In the following we present two examples of how evolution dynamics can be treated methodologically in this manner. 
The first deals with cascades of creation and destruction in evolutionary systems. This is done in section \ref{CoCD} where we 
also show how to extend the analysis of earlier work\cite{hanel05,hanel07}. The second example deals with the formation process in section \ref{OF}.

\section{Creative and destructive cascades} 
\label{CoCD}

We have analyzed systems
where pairs of elements exist that can be combined to form a third element\cite{hanel05,hanel07}. 
Such systems are sometimes called catalytic networks.
We have demonstrated the existence of both creative and destructive cascades (algorithmic phase transitions).
We first review the results and then extend them to the situation where $n$ elements can be combined to form a product-element. 
Note, that this model is a mathematically tractable variant of the so called  bit-string model 
of biological evolution\cite{origin}. 

Suppose there is a maximal diversity of elements in the system $d$ that can possibly exist and write the relative frequencies
$x_i$ (how much there is of $i$) of the elements $i$ as the $d$ dimensional vector $x=(x_1,x_2,\dots,x_d)$.
This vector $x$ describes the state of the system. 
The dynamics of $x$ is governed by the famous {\em catalytic network equation}
\begin{equation}
\frac{d  }{d t} x_i = \sum\limits_{j,k}\alpha_{ijk} x_j x_k - x_i \sum\limits_{ljk} \alpha_{ljk} x_j x_k  \quad, 
\label{model}
\end{equation}
where the second term ensures normalization of $x$. The tensor elements $\alpha_{ijk}$ serve as the 'rule table', telling which combination of two elements $j$ and $k$ can produce a third (new, different) element $i$. The rule table element $\alpha_{ijk}$ is the rate at which element $i$ can get produced, given the elements  $j$ and $k$ are abundant at their respective 
concentrations $x_j$ and $x_k$. We call a pair of elements $(j,k)$ {\em productive} when there exists an $i$ such that 
$\alpha_{ijk}>0$ is strictly larger than zero.
Special cases of Eq.\,(\ref{model}) are the Lotka Volterra replicators\cite{lotka}, the hypercycle\cite{hyper}, or the Turing gas\cite{turing}. Equation\,(\ref{model}) has been analyzed numerically\cite{farmer1,numerical}, however for rather moderate system sizes. 

Many goods and services require more than two inputs. For instance, a car-engine consists of many parts that have to be put together to form a functional entity, or the production of wine requires barrels, grapes, bottles, human labor and appropriate  quantities of sun and rain. Therefore we generalize the catalytic network equation to processes, where $n$ elements instead of two elements are necessary to form a new one.
To do so, 
%%%%%%%%%%%%%%%%%%%%%%%%%%%%%%%%
% RUDI: 13 to 14: 1.) semicollon inserted
%%%%%%%%%%%%%%%%%%%%%%%%%%%%%%%%
we have to pass from considering index pairs $(j,k)$ like in Eq.\,(\ref{model}), to $n$ dimensional index vectors 
${\bf j}\equiv(j_1,\dots,j_n)$ to indicate the necessary  $n$ substrate elements.
Therefore the rule table becomes a tensor $\alpha_{i {\bf j}}\equiv\alpha_{i,j_1,j_2,\dots,j_n}$ depending on the product $i$ and $n$ substrate-elements $j_m$, where $m=1,2,\dots,n$. To model the effect of the availability of elements on the production 
we also have to pass from the frequency pair $x_j x_k$ in Eq.\,(\ref{model}), to
the product of the $n$ frequencies $x_{j_1}x_{j_2}\dots x_{j_n}$. 
The notion of the productive pair naturally extends to the notion of {\em productive sets} such that a set of elements ${\bf j}$ is productive when there exists an $i$ such that $\alpha_{i{\bf j}}>0$.
Replacing this appropriately into Eq.\,(\ref{model}) gives us the {\em generalized catalytic network equation}.
Remarkably, this generalization does not  alter the arguments that allow us to analyze the creative and destructive cascades of production we are interested in. 

\subsection{Cascades of creation}

First of all, applying (i) from the previous section,
we reduce the dimensionality of the problem by counting the elements $x_i>0$, 
and define the relative diversity $a_t\in [0,1]$ of the system such that $a_t\,d$ is the number of existing elements at time $t$.  
We assume that the system has a growing mode only (tensor elements $\alpha\geq 0$, i.e. $\alpha=1$ but never 
negative). Entries $1$ in the {\em rule-table} tensor $\alpha$ indicate the {\em productive sets}. 
Suppose, applying (ii), we only know the number $r$ such that $r\,d$ is the number of such productive sets occurring in the system and that the entries $1$ and $0$ in $\alpha$ are distributed completely randomly. 
$r$ constitutes our constraining knowledge on $\alpha$.
The only variables characterizing our system therefore are $r$ and $n$.
Let us suppose -- as the initial condition -- that initially,  at time $t=0$, only a few elements 
$a_0\,d$ exist. This means that $a_0 d$ elements in the initial element vector are different from zero. 
We are interested in the asymptotic value ($t\to \infty$) of the relative diversity $a_{\infty}$.

%%%%%%%%%%%%%%%%%%%%%%%%%%%%%%%%
% RUDI: 13 to 14: 2)
%%%%%%%%%%%%%%%%%%%%%%%%%%%%%%%%
%The fraction of productive sets of elements in the system is given by
The fraction of sets of $n$ elements in the system which are productive is given by
%%%%%%%%%%%%%%%%%%%%%%%%%%%%%%%%
$r\,d$ divided by the binomial coefficient of $d$ over $n$ elements $\binomial{d}{n}$. In a given set of size $a_t d$ there are 
$\binomial{a_t d}{n}$ 
%%%%%%%%%%%%%%%%%%%%%%%%%%%%%%%%
% RUDI: 13 to 14: 3)
%%%%%%%%%%%%%%%%%%%%%%%%%%%%%%%%
%productive sets 
sets consisting of $n$ elements
%of elements 
%%%%%%%%%%%%%%%%%%%%%%%%%%%%%%%%
to be found on average. Due to the assumed randomness of the distribution of productive sets of elements the number of produced elements is
\begin{equation}
rd\binomial{a_t d}{n}\binomial{d}{n}^{-1}\approx\: r\,d\,a_t^n\quad.
\end{equation} 
Yet, about $r\,d\,a_{t-1}^n$ of those elements already have been produced in the preceding time step $t-1$ and are certainly contained in $a_t$ already. They consequently have to be subtracted as already existing. Due to the randomness of the production of elements only a fraction $1-a_t$ of the remaining elements will not be present in $a_t$ and truly be newly created. Putting all this together yields 
\begin{equation}
 a_{t+1}=a_{t}+\Delta a_{t} \quad , \quad
 \Delta a_{t}=r\left(1-a_{t}\right)
 \left(a_{t}^{n}-a_{t-1}^{n}\right)   .
\label{update}
\end{equation} 
at timestep $t=0$ we have to set $a_{-1}\equiv0$, by convention since $a_0$ is the initial condition. 

\begin{figure}
\begin{tabular}{cc}
\includegraphics[angle=270, width=7cm]{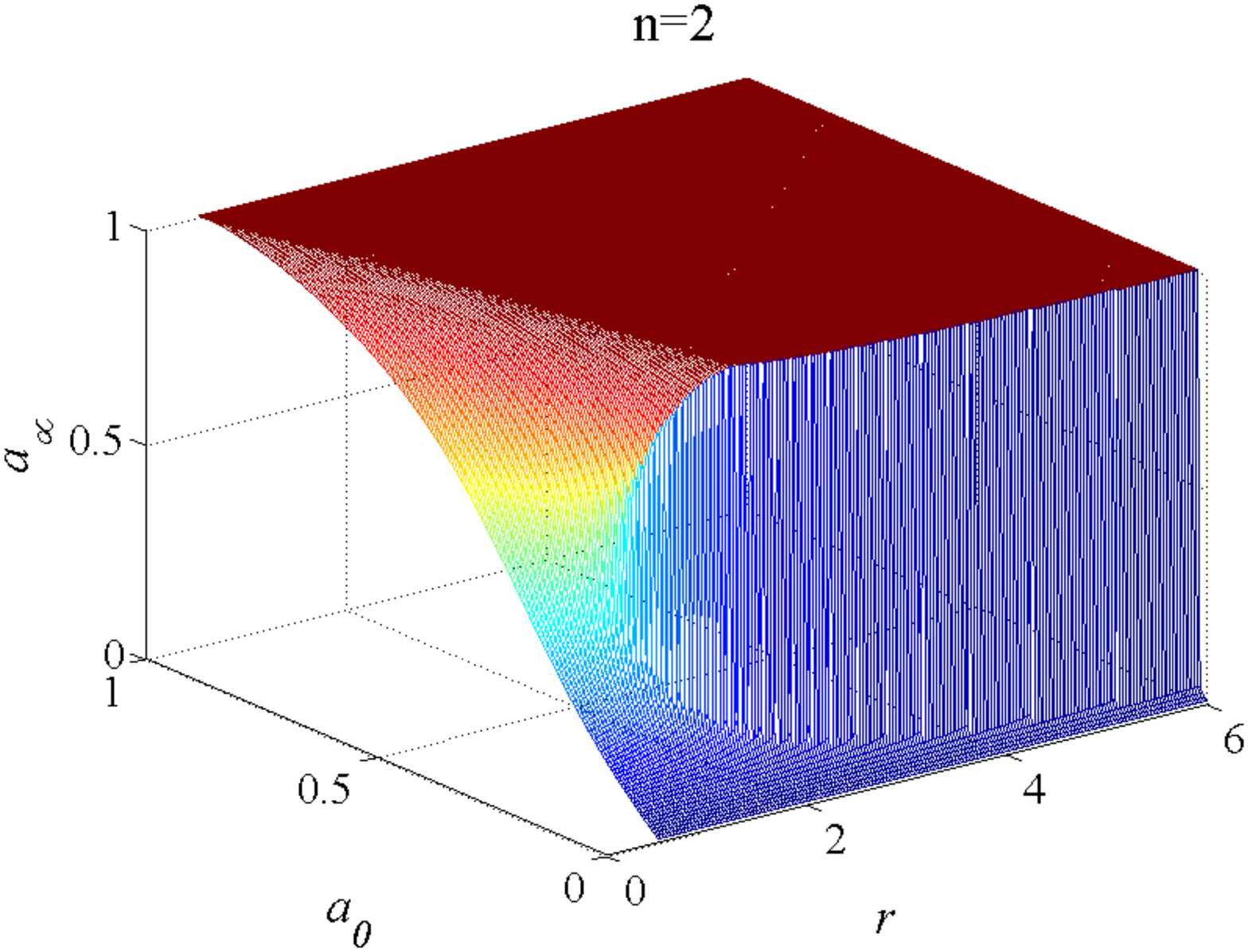} &
\includegraphics[angle=270, width=7cm]{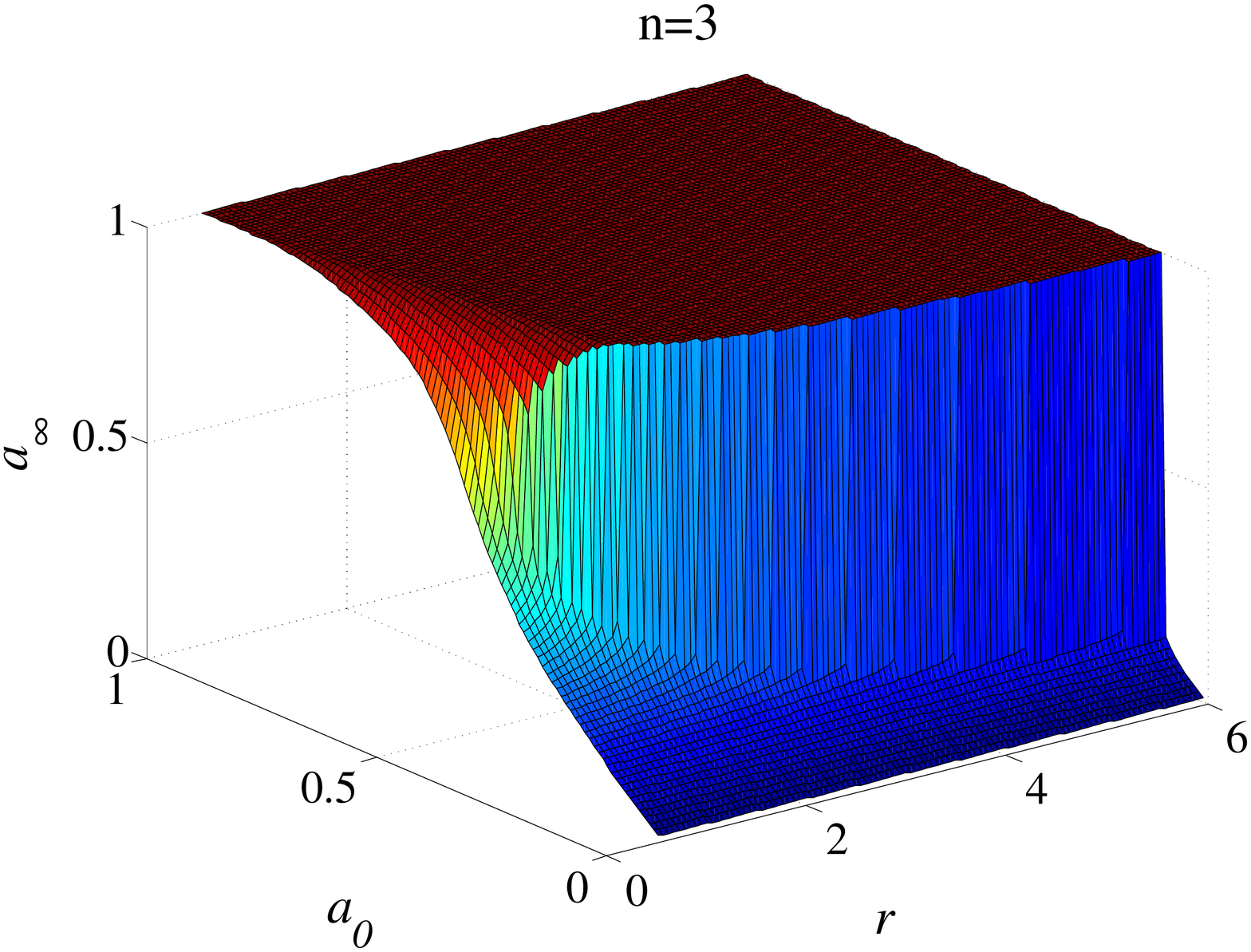} 
\end{tabular}
\caption{Phase diagram of the creative dynamics over the $r$-$a_0$ plane for $n=2$ and $n=3$.
The diagram shows clearly how the system starts to exhibit a phase transition for $r\sim 2$ in the $n=2$ case and
how the critical initial complexity $a_0$ where the phase transition occurs gets smaller for increasing values of $r$.
The phase transitions occurs for larger values of $a_0$ in the $n=3$ case. 
}
\label{fig1}      
\end{figure}

These equations are exactly solvable in the long-time limit. For this end 
define,  $c_{t}\equiv \Delta a_{t+1}/ \Delta a_{t}$, and look at the asymptotic behavior,  
$c \equiv \lim_{t\to\infty}c_t$.
From Eq. (\ref{update})  we get 
\begin{equation}
 c=nr\left(1-a_{\infty}\right)a_{\infty}^{n-1}  \quad. 
 \label{ansatz_a}
\end{equation} 
On the other hand we can estimate $a_{\infty}$ asymptotically by
\begin{equation}
 a_{\infty}= a_0 \sum_{t=0}^{\infty}c^t = \frac{a_0}{1-c} \quad .
\label{ansatz_b}
\end{equation} 
By inserting  Eq.\,(\ref{ansatz_a}) into Eq.\,(\ref{ansatz_b})
one gets a $N+1$'th order equation
\begin{equation}
a_\infty-a_0= nr(1-a_\infty)a_{\infty}^n\quad,
\label{asymptotic_phase_equ}
\end{equation} 
whose solutions are the solution to the problem. Most remarkably these solutions 
are mathematically identical to the description of real gases, i.e. Van der Waals gases, independent of the parameter $n$. 
One can make the relation to the  Van der Waals gas explicit by 
defining, $V \equiv a_{\infty}$, $T  \equiv a_0(nra_\infty^{n+1})^{-1}$, and $P \equiv a_\infty^{-1}+(nra_\infty^{n+1})^{-1}$. 
This allows to generically map Eq.\,(\ref{asymptotic_phase_equ}) 
onto the famous equation
\begin{equation}
\left(P-\frac{1}{V^2}\right)V=T\quad, 
\label{vdwaals}
\end{equation}   
which is exactly a Van der Waals gas of point-particles with constant  internal pressure.
The meaning of 'pressure' and 'temperature' in our context is intuitively clear. 
As comparable to real gases our $n$-ary assembly system shows a phase 
transition phenomenon which are of the same universality class independent of the choice of $n$.
The corresponding phase diagram,  
as a function of the model parameter $r$ and the initial 
condition $a_0$ is shown in Fig. \ref{fig1} for the choice $n=2$ and $3$. 
Note that assembly systems with larger values of $n$ undergo the phase transition at larger values
of $a_0$ for fixed $r$ so that larger sets of random entities are required to trigger self-sustaining creative cascades.

The update-equations (\ref{update}) are scale invariant. 
The solutions of the problem in terms of relative diversities do not explicitly depend on the upper 
bound of diversity $d$  and the results therefore apply also in the limit $d\to\infty$.
One can think of open systems which can not be described as a limit of bounded systems.
Such systems can be treated in a similar fashion, however this task is of considerably higher mathematical complexity. 
Work dealing with this problem is in preparation. 

\subsection{Cascades of destruction}

In the dynamics studied so far, diversity can only increase due to the positivity of the  
elements in $\alpha$. It is important to note that in this setting 
the phase transition can not be crossed in the backward direction. 
This is because of two reasons. 
First, the system {\em forgets} its initial condition $a_0$ once it
has reached the (almost) fully populated state. This means that after everything has been 
produced one can not lower the initial set size any more. In terms of  the Van der Waals gas equation
analogy we can {\em not} lower the 'temperature'.
Second, if $r$ and $n$ are homogeneous characteristics of the system then it is also impossible to manipulate 
the 'pressure' of the system and we remain in the fully populated phase for ever.
The natural question thus arises what happens to the dynamics if one 
randomly kills a fraction of elements in the fully (or almost fully) populated phase. 
In the case that an element $k$ gets produced by a single productive set ${\bf j}$ and one of the members
of the $j_m$ gets killed, $k$ can not be produced any longer. We call the random removal of $i$ 
a {\em primary defect}, the result -- here the stopped production of $k$ -- 
is a {\em secondary defect}. 
The question is whether there exist critical values for $r$ and the primary defect density $\delta_0$, 
such that cascading defects will occur. 

\begin{figure}[t]
\begin{tabular}{cc}
\includegraphics[angle=270, width=7cm]{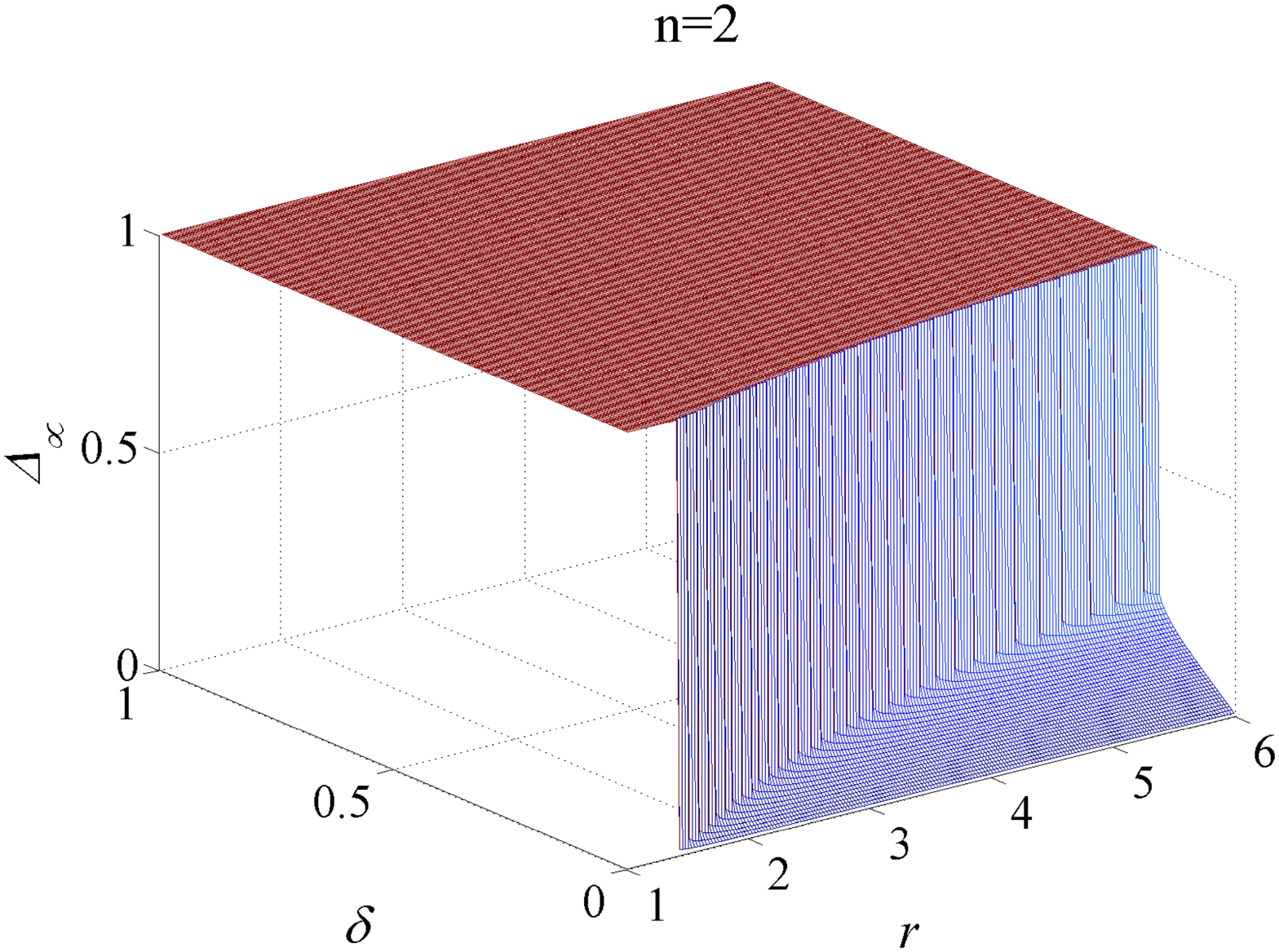} &
\includegraphics[angle=270, width=7cm]{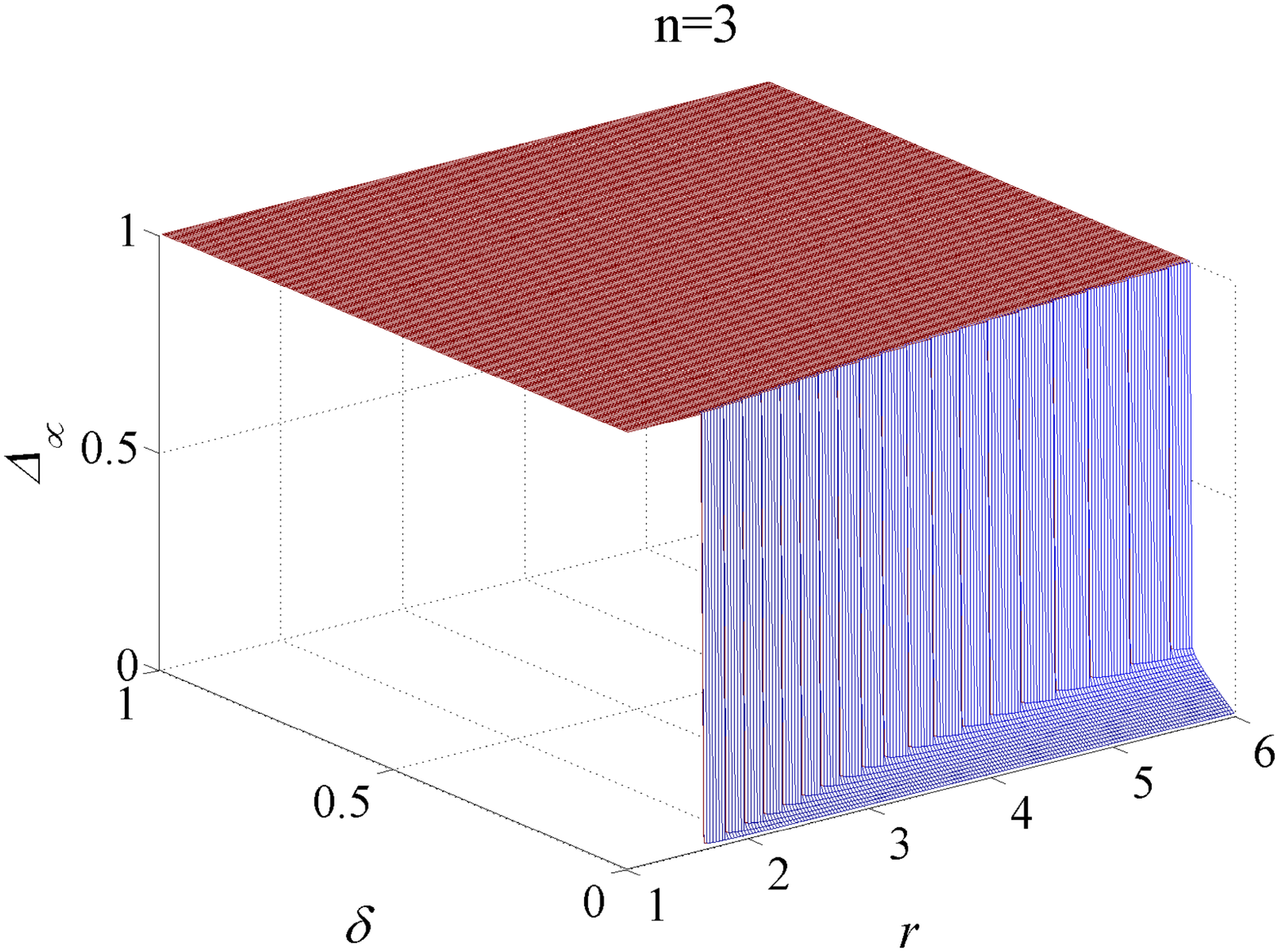} 
\end{tabular}
\caption{Phase diagram of the destructive dynamics over the $r$-$\delta_0$ plane for $n=2$ and $n=3$.
We see that relative small numbers of initial defects suffice for the whole system to break down.
The culminated defects $\Delta_\infty$ first raise linearly dominated by the number of primary defects $\delta_0$. 
At the critical of primary defects the system breaks down almost without precursor in the number of culminated defects.
We may also see how the $n=3$ case is considerably more sensible to defects than the $n=2$ case. 
}
\label{fig2}      
\end{figure}

As before one may approach this question iteratively, by asking how many secondary defects will be 
caused by an initial set of $\delta_0 d$ randomly removed elements in the  fully populated phase. 
The possibility for a secondary defect happening to element $k$ requires that {\em all}  
productive sets, which can produce $k$, have to be destroyed, i.e. 
at least one element $j_m$ of the productive set has to be eliminated for each productive set {\bf j}. 
On average there are $r$ such productive sets for each element $i$.
Let us introduce the first defect. 
This defect will on average affect $nr$ productive sets in the system, i.e.,  
there will be $nr$ elements that lose one way of being produced. 
This is so, since there are $d\,r$ productive sets and there are $n\,d\,r$ indices referring to an 
element involved in denoting the productive set. Consequently there are $n\,r$ indices on average per element. 
To look how adding primary defects effects the system and the number of elements that have lost one, two, and more ways of getting produced iteratively requires some 'book-keeping'. 
However, from here on we basically merely have to adapt a derivation presented in earlier work\cite{hanel07} in a straight forward fashion. Since the mathematics is not difficult but lengthy we directly jump to the result
\begin{equation}
 \delta_1=\gamma(r)f_{r\,n}(\delta_0)\delta_0^{r}\quad,
 \label{sec_defects_d}
\end{equation}   
with 
\begin{equation}
 \begin{array}{lll}
     \gamma(r)&=&\frac{1}{r}\frac{(nr)^r}{\sqrt{2\pi}}(r-1)^{\frac{1}{2}-r}e^{r-1}\\&&\\
  f_{r\,n}(\delta_0)&=&\sum_{k=0}^{\infty}\frac{1}{k!}\frac{r}{r+k}(-nr\delta_0)^k\quad. 
 \end{array}
 \label{sec_defects_ad_d}
\end{equation}   
The result remains to be iterated from secondary to tertiary and higher order defects. 
To do so we think of collecting the primary and secondary defects together and
assume that we would start with a new primary defect set of relative size $\delta_0'=\delta_0+\delta_1$. 
The tertiary defects therefore gets estimated by 
$\delta_2=\delta_1'-\delta_1$, 
where $\delta_1'$ are the secondary defects associated with $\delta_0'$. This leads to the recursive scheme, 
\begin{equation}
\Delta_m\equiv\sum_{k=0}^m\delta_k \quad,\qquad 
\delta_{m+1} = \gamma(r)f_{r\,n}(\Delta_m)\Delta_m^r -\Delta_m + \delta_0\quad,
\label{defect_recrel_A}
\end{equation}
where $\Delta_t$ are the culminated defects after $t$ timesteps.
The result in terms of a phase diagram of the iteration scheme is 
given in Fig. \ref{fig2} for $n=2$ and $3$. 
The asymptotic 
%%%%%%%%%%%%%%%%%%%%%%%%%%%%%%%%
% RUDI: 13 to 14: 4) inserted ...
%%%%%%%%%%%%%%%%%%%%%%%%%%%%%%%%
culminated
%%%%%%%%%%%%%%%%%%%%%%%%%%%%%%%%
total defect size $\Delta_{\infty }$ (for $t\to \infty$) is shown as a function of the 
parameters $r$ and the initial defect density $\delta_0$. As before 
a clear phase transition is visible, meaning that at a fixed value of $r$ there exists a critical 
number of initial defects at which the system will experience a catastrophic decline of diversity. 

It is interesting that for the complete destruction of diversity (plateau in Fig. \ref{fig2}) not very large values 
of $\delta_0$ are necessary. We also want to stress that assembly systems with larger values of $n$ 
collapse at smaller values of $\delta_0$. They are more sensitive to the exposure to initial defects.

In a more realistic scenario one can assume that (exogenous) random processes are introducing 
new elements at a given rate into a system just as other (exogenous) random processes 
may introduce defects. We have shown\cite{hanel07} how, being equipped
with the update equations for the creative and the destructive cases Eqs. (\ref{update}) and (\ref{sec_defects_d}), 
these update equations can be coupled to study the combined dynamics driven by exogenous random events.
This allows to take a look at how diversity of systems may evolve over time, driven by the spontaneous creation and destruction processes, which may reflect exogenous influences.
There we have also shown that the increment distribution of $\Delta a_t\equiv a_t-a_{t-1}$ is power-law distributed, when the
system is prepared critically, i.e. at the onset of the creative phase-transition. 

We may conclude from our analysis of generalized catalytic systems that average number of elements required to build a new element of the system does not change the overall picture previously drawn\cite{hanel05,hanel07} although larger $n$ make it more difficult to reach the creative phase transition in the low diversity phase and easier to reach the destructive phase-transition in the high diversity phase.
The limit $n\to 1$ is comparable with the situation of the opinion formation problem under the unanimity-rule\cite{Lambiotte06} for an average connectivity $r$ of each node. 
Since the stability of biological networks are often analyzed under the unanimity-rule governing trophic dependencies\cite{ht_foodweb, dunne02_foodweb_1, sole01_foodweb}, 
where a species can survive as long as it has one remaining trophic link,
it is interesting to see that this is the most stable model-assumption one can pose ($n=1$);
this is to say trophic-dependencies-only is a {\em friendly} scenario. 
If the survival of species are also depending on other than trophic factors this can only result in an increase 
of the parameter $n$ to values larger than one. Thus biological networks can be considerably less stable 
than the analysis, based on trophic dependencies only,  would suggest.

\section{Formation dynamics} 
\label{OF}

\begin{figure}[b]
\begin{tabular} {c}
\includegraphics[height=4cm] {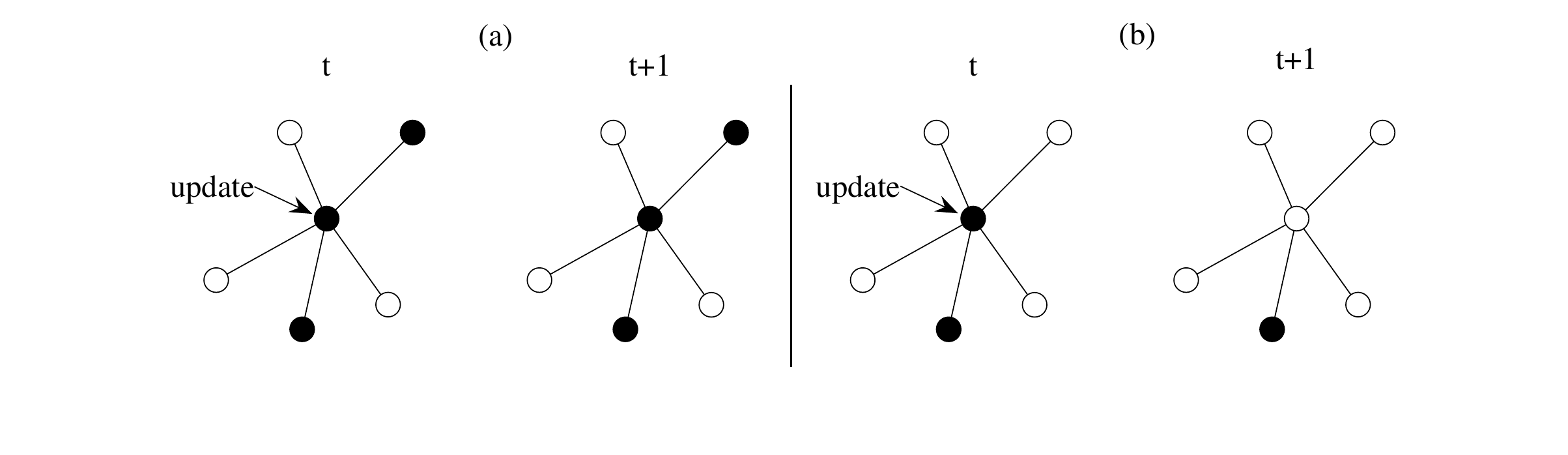}
\end{tabular}
\caption{A pictorial description of the update process with $h=0.8$. The node marked with an arrow gets updated, it has five neighbors for which we show two possible configurations of internal states. (a) Three out of five neighbors are in a different state, the node stays unchanged. (b) Four out of five neighbors are in a different state leading to a change in the internal state of the updated node.}
\label{upd} 
\end{figure}
 
So far we have been interested in creation and destruction processes on catalytic sets. 
We now turn to the analysis of the formation processes in evolutionary dynamics. 
It was shown that the formation problem of systems of evolution can be mapped onto 
an opinion-formation problem\cite{Lambiotte06}.
As mentioned in the introduction, elements in this scenario exist in two different states 0/1 (e.g. existing/not existing) and each element is connected with a set of other elements through a (fixed) interaction matrix. An element can change its state depending on the states of the elements it is connected to. For example, some type of molecule may change its state from non existing to existing when sufficiently many other molecules, from which it can be produced in chemical reactions, also exist - or vice versa. 
Another example, which can be understood as an opinion-formation problem, is the ongoing debate about the standard for high definition DVDs (Blu-ray against HD-DVD), paralleling the historical Betamax vs. VHS conflict. 
We assume that one is more likely to favor one alternative if this alternative is also favored in one's local neighborhood. If all of my friends would have shared their tapes in the VHS standard (because a certain branch of movies is only released in this format), it is more likely that I would have dumped my Betamax video recorder and used VHS from then on. It is also possible that two different opinions can co-exist, e.g. Mac and Windows users. In the next section we will propose a model capable of  showing these phenomena. A node will change its internal state if the fraction of neighbors holding the opposite opinion (state) is above a predefined threshold $0.5<h\leq1$. These opinion formation models are sometimes called threshold voter models\cite{Liggett99, Watts02}.

\subsection{The model}

We represent an agent $i$ as a node in a network composed of  $N$ agents. The internal state of the node corresponds to the agent's choice, e.g. which technology he uses in every-day situations.
For simplicity we break down each decision into binary choices, yes/no, VHS/Betamax, MS Windows/Mac, etc.
Connected nodes influence each other in the decision-making process, i.e. they tend to persuade their 
neighbors to adopt their own choice.  
The decision-making process consists of four steps
(see Fig.\ref{upd}): 
\begin{itemize}
\item  Pick a node $i$ randomly.
\item  Check the internal state of each of $i$'s neighbors. 
\item  If the fraction of states holding opinion 1(0) is above the threshold $h$, $i$ adopts opinion 1(0).
\item  If the update threshold is \emph{not} exceeded, $i$ keeps its initial state.
\end{itemize} 

\begin{figure}
\begin{tabular} {ccc}
\includegraphics[angle=270, width=5.28cm]{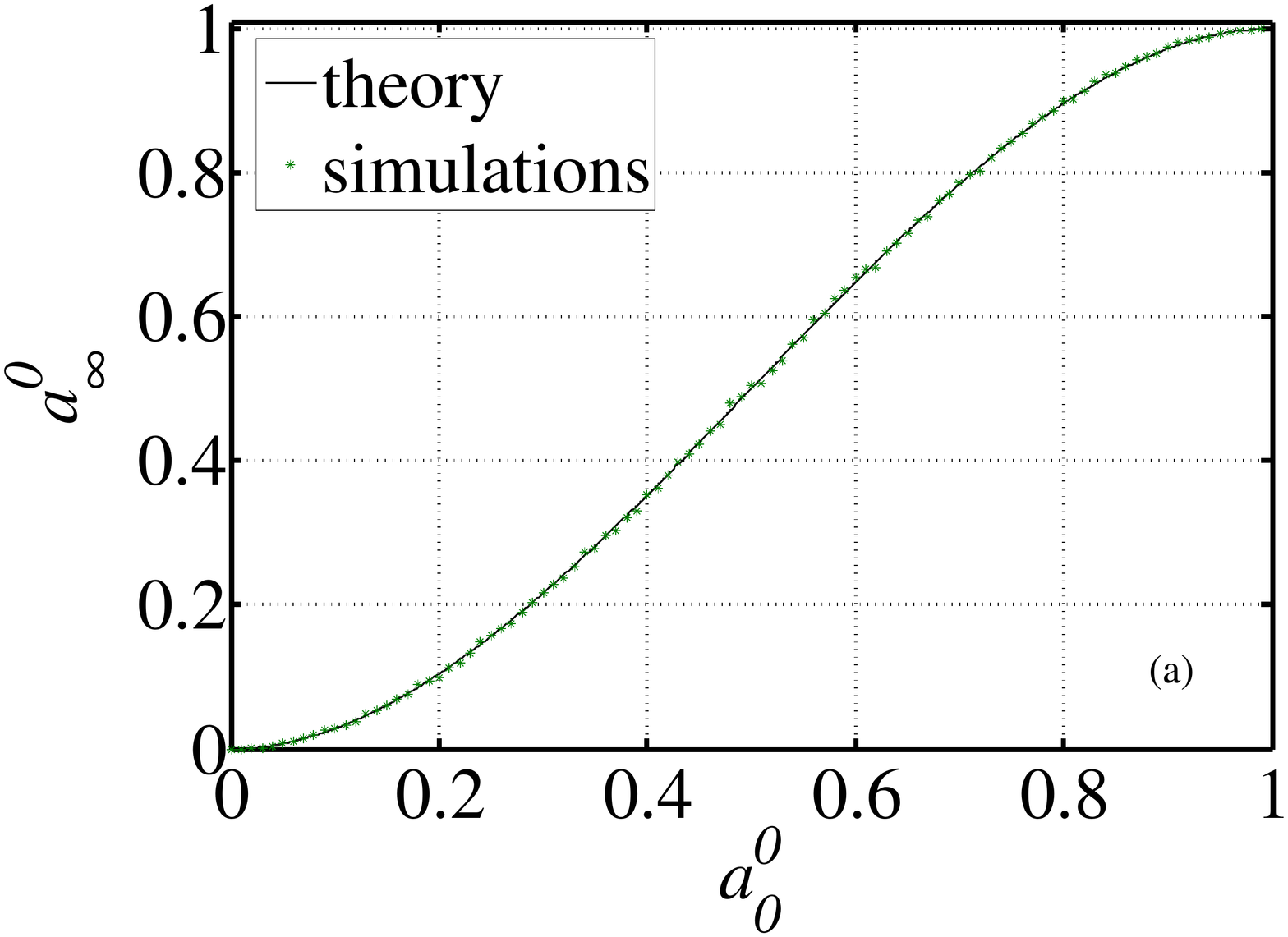}    &
\includegraphics[angle=270, width=5.28cm]{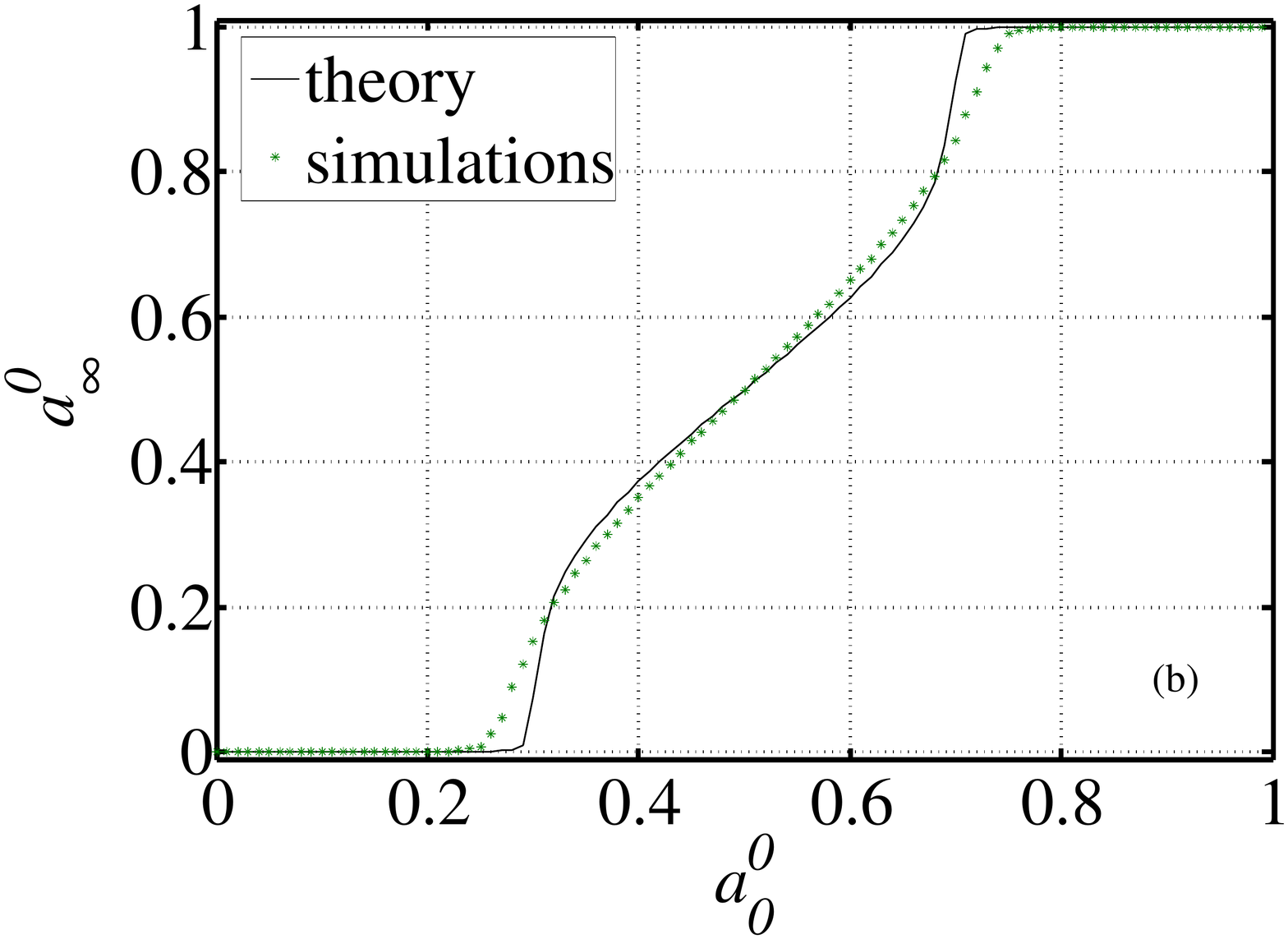}   &
\includegraphics[angle=270, width=5.28cm]{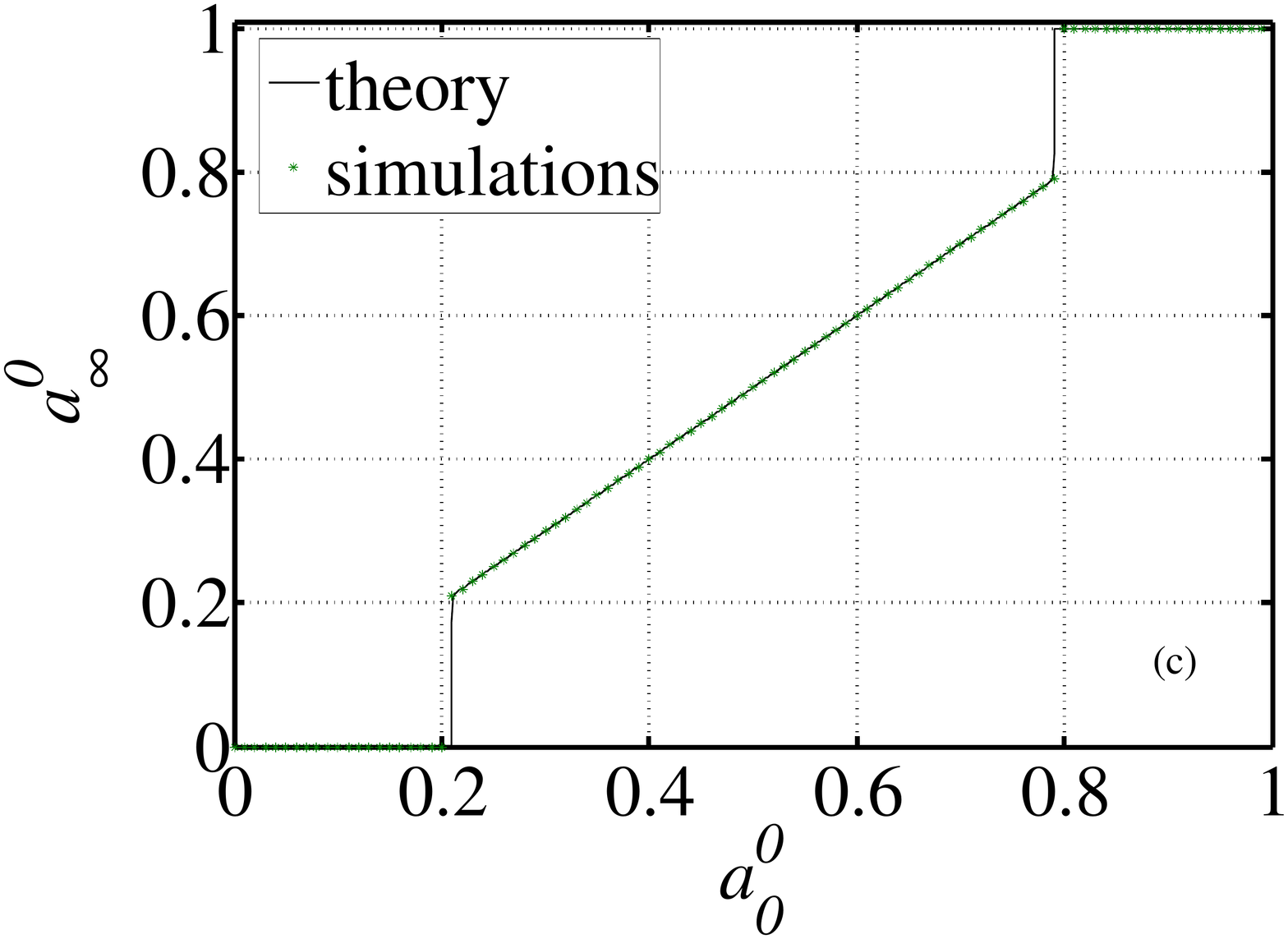}   
\end{tabular}
\caption{The order parameter $a_{\infty }^0$ as a function of the initial population sizes, $a_0^0$,   
for $N=10^4$, $h=0.8$. (a) $k=2$ for all nodes (unanimity regime),  (b) random network with $\bar{k}=10$ (intermediate regime) and (c) random network with $\bar{k}=9 \times10^{3}$. With increasing $\bar{k}$ the transition between the phase of coexistence and consensus becomes sharper.}
\label{Pop}
\end{figure}

We assume that the only information available about the topology of the network is the probability for any two nodes to be connected. Our ignorance thus tells us, following model assumption (ii), to model the interaction matrix
as a random network\cite{ER} where  $N$ nodes are randomly linked with $L$ links (self-interactions are forbidden). 
Of course any other network topology could be used, however at the cost of more mathematical trouble\cite{abe}. 
If two nodes $i$ and $j$ are connected, we set the corresponding entries $(i,j)$ and $(j,i)$ to 1, if two nodes do not interact the entry would be set to 0. For sufficiently large systems the crucial parameter is the average connectivity $\bar k = L/N$ and our results are independent of the actual number of agents.   
The update threshold 
necessary for a node's change of opinion, 
$h$, has to be higher than $0.5$ in order 
to be meaningful in the above sense. The update is carried out random sequentially.
In a network containing $N$ agents at any time $t$, there are $A_t^0$ nodes 
with choice 0 and $A_t^1$ nodes with choice 1. 
We again use dimensionless variables and define the fraction of nodes in state 0/1 as $a_t^{0/1}=A_t^{0/1}/N$.
The role of the $a_t^{0/1}$ is comparable to the role of the relative diversity $a_t$ in the previous section, and reduce
the complexity of the problem according to model assumption (i). 
If the sequence of updates is complete we increase $t$ by 1 and start the next random sequence of updates. 
We are interested in the order parameter $\lim_{t \rightarrow \infty} a_t^{0/1} = a_{\infty}^{0/1}$.

\subsection{Theory and simulations}

We now derive a master equation for the evolution of this system by adopting the same philosophy as before. First, let us outline the main difference with respect to the previous scenario. In the creative dynamics we incorporated updates in one direction, e.g. from not-produced $\to$ produced. In the formation dynamics the updates are required to take place in both directions. Thus our aim is to derive the transition probabilities $\Delta_t^{0 \rightarrow 1}$ and $\Delta_t^{1 \rightarrow 0}$ for both processes, an update from 0 to 1 and vice versa with combinatorial arguments\cite{hanel05}. 
We start at $t=0$ where we have a fraction of $a_0^0$ nodes in state 0. The probability that at time $t$ 
an update from 0 to 1 occurs  is  denoted by $\Delta_0^{0 \rightarrow 1}$. 
This transition rate is the sum over the probabilities for all configurations of neighbors where at least a 
fraction of $h$ neighbors is in state 1 and we simultaneously have picked a node initially in state 0, 
\begin{equation}
   \Delta_0^{0 \rightarrow 1}=a_0^0 \sum^{\bar{k}}_{i=\lceil \bar{k} h\rceil} {\bar{k} \choose i} \left( 1-a_0^0 \right)^i \left( a_0^0 \right)^{\bar{k}-i} 
  \quad, 
  \label{prob}
\end{equation}
where $\lceil . \rceil$ denotes the ceiling function, the nearest integer being greater or equal. 
To derive $\Delta_0^{1 \rightarrow 0}$ one just has to exchange 0 and 1 in  Eq.(\ref{prob}). 
The equation governing the system's evolution from $t=0$ to $t=1$ thus is given by
\begin{equation}
  a_1^0=a_0^0+\Delta_0^{1 \rightarrow 0}-\Delta_0^{0 \rightarrow 1}  \quad. 
\label{master1}
\end{equation}
Let us examine some special cases now.

\subsubsection{Unanimity regime}

If $\bar{k}$ and the update threshold $h$ are chosen such that $\lceil\bar{k} h\rceil=\bar{k}$ holds, the evolution halts after one iteration. We call this the unanimity regime. Here the update rule boils down to the unanimity rule, i.e. all linked nodes have to be in the same internal state to allow an update. Let the following account for an argument: 
Consider the second 
%%%%%%%%%%%%%%%%%%%%%%%%%%%%%%%%
% RUDI: 13 to 14: 6)
%%%%%%%%%%%%%%%%%%%%%%%%%%%%%%%%
%interaction. 
iteration. 
%%%%%%%%%%%%%%%%%%%%%%%%%%%%%%%%
Assume that no consensus has been reached and we thus can find two neighboring nodes favoring different choices, say agent $i$ is in state 0, $j$ holds state 1. $i$ could be updated if each of his other neighbors is in state 1. But if this was the case, $i$ could \emph{not} be in 0. Either the update to 1 would have occurred in the first iteration, or there is an agent $k$ in $i$'s neighborhood  which also holds state 0 and can not be updated due to his connectedness with $i$. So, as one can easily check by inspection, the dynamics of the system freezes after the first iteration.
For the special case $k=2$ (constant) we find that $a_{\infty}^{0}=a_1^0$. 
Inserting this in Eq.(\ref{master1}) yields 
 \begin{equation}
   a_{\infty}^{0}=3(a_0^0)^2-2(a_0^0)^3 \quad. 
 \label{predk2}
 \end{equation}
Results of a simulation on a regular 1D circle network with $N=10^4$ confirm the theoretical prediction of Eq.(\ref{predk2}), see Fig.\ref{Pop}(a).

\subsubsection{High connectivities}

For higher values of $\bar{k}$ and $h$ the gridlock scenario described above does not appear. In subsequent iterations we still observe updates. However, after waiting sufficiently long the evolution converges to a halt.
As the fraction of nodes in one state increases, this raises the probability that the respective fraction of nodes will grow even further (as the words spread amongst the agents). This may force the system into consensus. Let us consider arbitrary time steps now.
To derive $\Delta_t^{0 \rightarrow 1}$ we look at the number of updates being possible with respect to the starting conditions, and we subtract the probabilities that these have already occurred at former time steps. This directly leads to 
\begin{equation}
\Delta_t^{0 \rightarrow 1}=a_0^0
\left(\sum^{\bar{k}}_{i=\lceil \bar{k} h\rceil} {\bar{k} \choose i} \left( 1-a_t^0 \right)^i \left( a_t^0 \right)^{\bar{k}-i} - \sum^{\bar{k}}_{i=\lceil \bar{k} h\rceil} {\bar{k} \choose i} \left( 1-a_{t-1}^0 \right)^i \left( a_{t-1}^0 \right)^{\bar{k}-i} \right) 
 \quad. 
 \label{prob2}
\end{equation}
Without being able to offer any more surprises we can finally state the master equation for arbitrary times,
\begin{equation}
a_{t+1}^0=a_t^0+\Delta_t^{1 \rightarrow 0}-\Delta_t^{0 \rightarrow 1}
\quad.
\label{master2}
\end{equation}
In Fig.\ref{Pop}(c) we test Eq.(\ref{master2}) against simulations for a system with $N=10^4$, $h=0.8$ and $\bar{k}=9 \times 10^3$ and find a formidable level of agreement. The plot allows us to identify three regimes. There are two regions of consensus with a sharply separated linear regime of coexistence.

For the case of a fully connected network we can state the asymptotic population sizes explicitly. If the initial populations are both smaller than the update threshold, i.e. $a_0^0<h$ and $1-a_0^0<h$, there will be no updates at all. Otherwise the system reaches consensus. So, for $k=N-1$ we find
\begin{equation}
a_{\infty}^0=a_0^0+\left(1-a_0^0\right) \Theta \left(a_0^0-h \right)-a_0^0 \Theta \left(1-a_0^0-h \right)
\quad,
\label{fc}
\end{equation}
where $\Theta\left(x\right)$ is the Heaviside step function. 
How can we understand the transition between this curve and the smooth solution found in Fig.\ref{Pop}(a)? 

\emph{Intermediate regime}.  
An interesting case appears when we leave the unanimity regime but keep the connectivity low compared to $N$. We have performed simulations with $N=10^4$, $h=0.8$ and $k=10$. Results are shown in Fig.\ref{Pop}(b) and compared to the theoretical prediction of Eq.(\ref{master2}). Although the agreement can be coined fair on a sunny day, it is not as good as in the previous cases.
The question arises what is missing in our description. In the iterative scheme we effectively calculate the relative numbers of updates without assuming any kind of correlations between the configurations at two different time steps. When we know that an update has taken place we have revealed some information about the neighborhood of the updated node. As an extremal case of this we have seen in the unanimity regime that after the first iteration no more updates can take place. In our analytical work we have not included this knowledge and assume the local neighborhoods to be maximally random with respect to the constraining population sizes. While this is a fair assumption for the vast majority of values for $\bar{k}$, there exists an intermediate regime where this is not the case. 
\begin{figure}
\begin{tabular}{cc}
\includegraphics[angle=270, width=7cm]{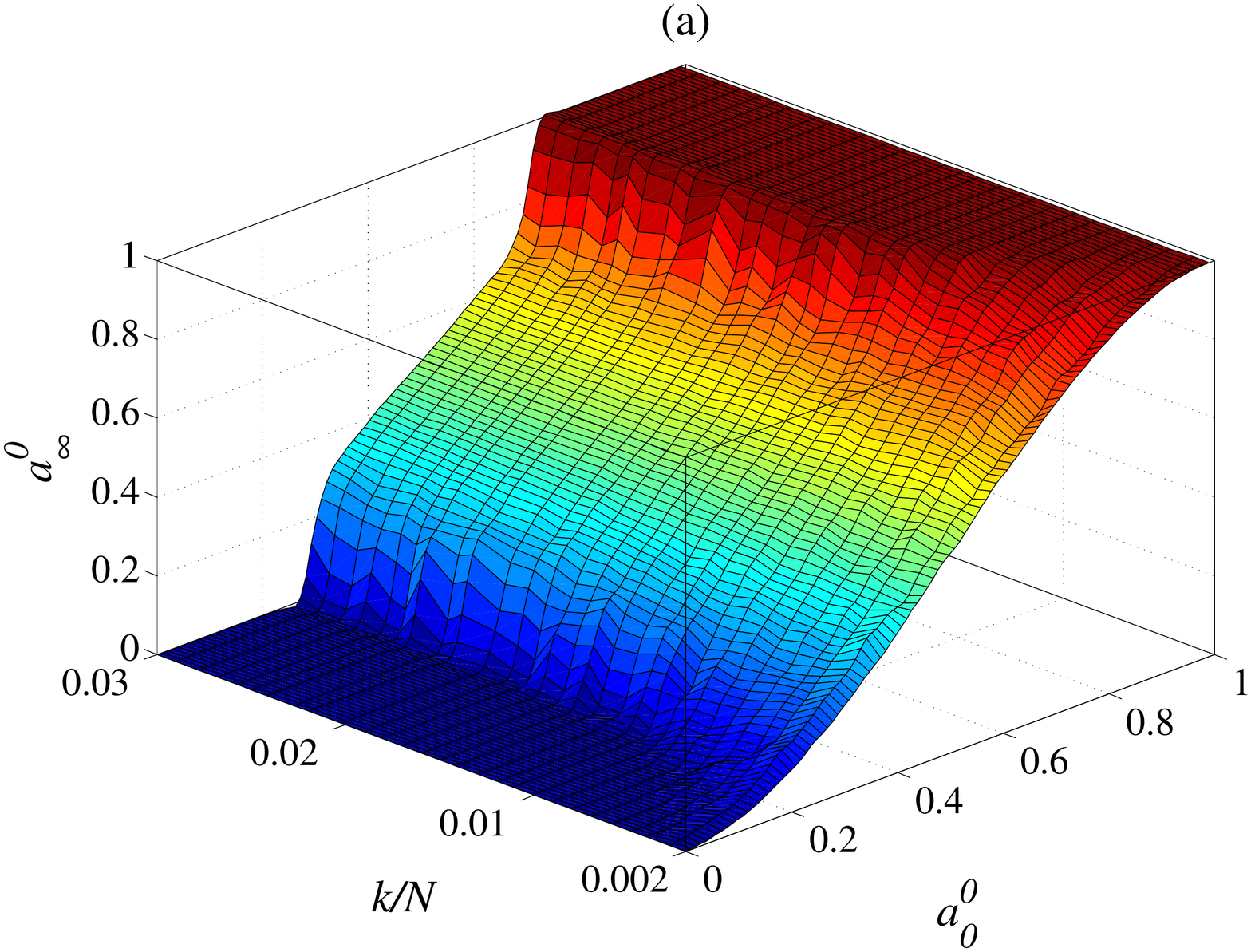} &
\includegraphics[angle=270, width=7cm]{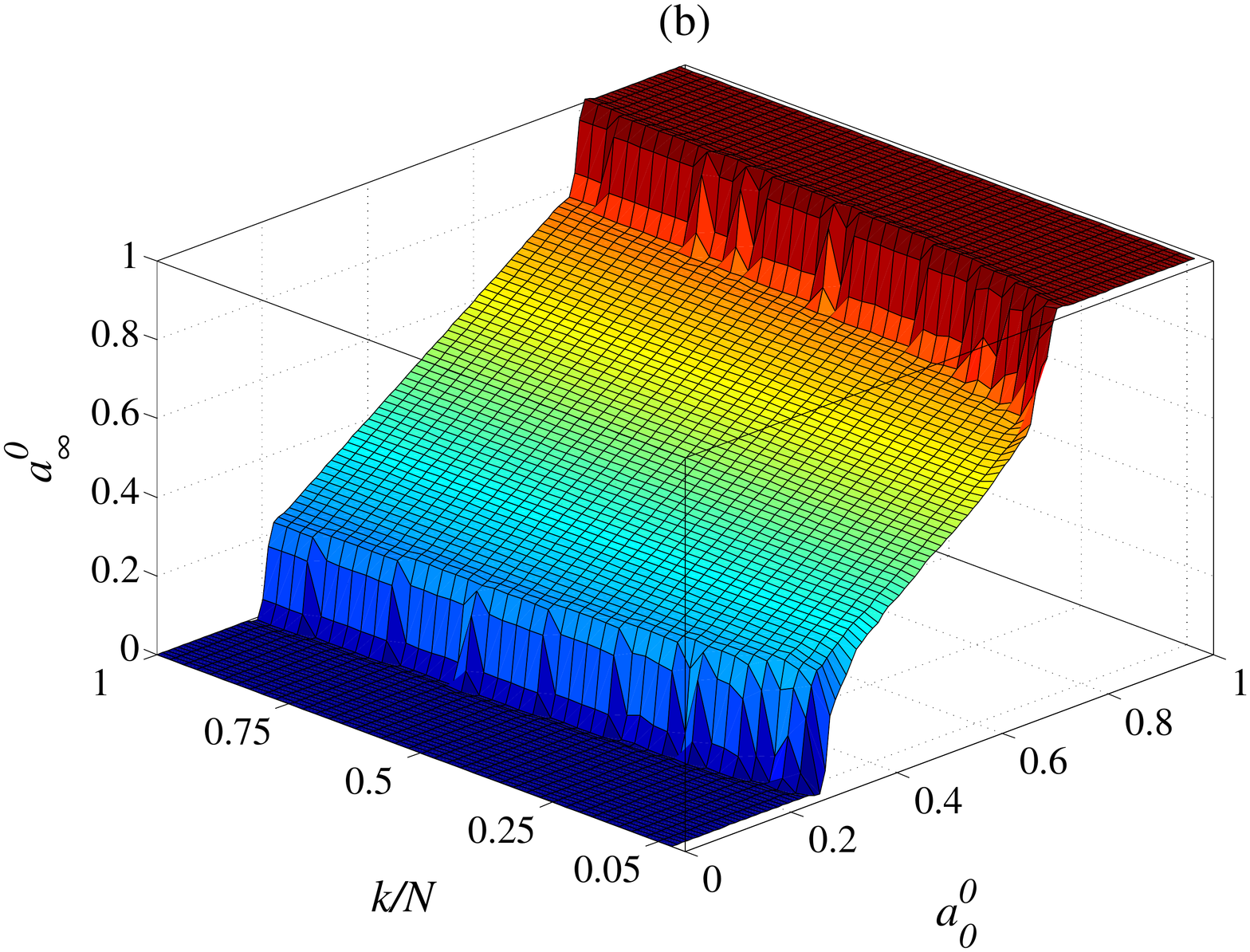} 
\end{tabular}
\caption{
$a_{\infty}^0$ as a function of the initial fraction size $a_0^0$ and connectedness $\bar{k}/N$. Simulations were performed on random networks with $N=10^3$ and $h=0.8$. Two symmetrical regions of consensus and a mixed phase in between are observed. (a) Phase diagram for the unanimity and the intermediate regime. With increasing connectedness the regions become more distinguishable. (b) For even higher values of the connectedness the dependence of the initial fraction size becomes linear in the mixed phase.
}
\label{phase}      
\end{figure}
The dynamics of the system is shown in the phase diagram, Fig.\ref{phase}. 
It illustrates the size of the respective regimes and their dependence on the 
parameters $a_0^0$ and connectedness $\bar k/N$.

For the voter-model it is known that the system will always end up in a consensus state. In contrast to this we also find stable solutions for coexistence. 
This sheds some light on the question 
%%%%%%%%%%%%%%%%%%%%%%%%%%%%%%%%
% RUDI: 13 to 14: 7)
%%%%%%%%%%%%%%%%%%%%%%%%%%%%%%%% 
%on which parameters it depends if a collective decision-making process 
of which parameters are controlling whether a collective decision-making process 
%%%%%%%%%%%%%%%%%%%%%%%%%%%%%%%%
ends in consens (like in the VHS-Betamax example) or in a coexistence (e.g. the Mac and Windows users).
 
This is possible due to two ingredients in the model presented here. First, our updates are carried out random sequentially. As a consequence the distribution of opinions among the nodes does not stay maximally random (w.r.t. initial conditions) and groups of agents sharing the same opinion appear. Second, the threshold $h$ can also be viewed as a level of tolerance in the sense of allowing a node to maintain its internal state even though a majority of the neighbors thinks different.

\section{Conclusions} 
\label{sec: conclusions}

Common ground in systems capable of evolution is that for the production of new species/goods/ideas etc. 
the existence of specific other items is a necessary precondition. 
In this context (auto)catalytic systems has become a cornerstone in understanding  evolutionary processes in various fields.
Notoriously high dimensionality of problems and poor knowledge on the exact form of interactions governing such systems complicate the situation.
We have shown in previous work how reducing the description of dynamics from the detailed level of the 
frequency of elements to the level of system diversity and how modeling interactions as random (when no better knowledge is available) paves a path towards the analytic treatability of systems  of evolution. 
We present two examples to make  clear the power of the arguments and the wide applicability of the method.
First, we applied the philosophy to massive events of creation and destruction in technological and biological networks. Here we extended and generalized earlier results on catalytic systems.
We find cascading events. Generalized catalytic systems can pass from low but critically large initial diversity to high diversity and they can break down from high to low diversity at a critical point of primary defects (removal of elements). 
The existence of this kind of phenomena in evolution is long known\cite{gould} in biology (e.g. the Cambrian explosion)  
and has been formulated in heuristic and qualitative terms in the economical context by J.A. Schumpeter. 
With our work we increase our ability to describe his 'creative gales of destruction' in a more quantitative framework.
Further we show that the new methodology allows to mathematically treat a variant of the threshold voter-model of opinion formation on random graphs. For fixed topology, we find distinct phases of mixed opinions and consensus. Stated in terms of systems of evolution the phases of consensus correspond to the phases of maximal and minimal diversity where either (almost) nothing or (almost) everything possible exists. The stable mixed opinions correspond to a phase where only a fraction of what possibly can be produced, exists in a stable way.
Whereas for the voter dynamics it is known that it always leads to a consensus state (i.e. only everything/nothing or one of two opinions/states), the model studied here allows the coexistence of two states.

\section*{Acknowledgements} 
Support from WWTF Life Science Grant LS 139 and Austrian Science Fund (FWF) projects P17621 and P19132.

\end{document}